\documentclass{appolb}
\usepackage{graphicx}
\usepackage{multirow}

\newcommand{\ra}{\rangle}
\newcommand{\la}{\langle}

\newcommand{\be}{\begin{equation}}
\newcommand{\ee}{\end{equation}}
\newcommand{\bea}{\begin{eqnarray}}
\newcommand{\eea}{\end{eqnarray}}
\newcommand{\ba}{\begin{eqnarray}}
\newcommand{\ea}{\end{eqnarray}}

\begin{document}
\title{Gluon confinement and the two IR solutions
\thanks{Light Cone 2012, 8/7/2012, Polish Academy of Sciences (Cracow). Thanks to Dan Zwanziger from New York University for his ideas and feedback on several topics. Support by spanish grants FPA2011-27853-01, FIS2008-01323, and the austrian FWF M1333-N16. }%
}
\author{Felipe J. Llanes-Estrada$^1$, Carlos Hidalgo-Duque$^2$, \\
and Richard Williams$^3$
\address{$^1$ Universidad Complutense de Madrid; speaker.\\
$^2$ IFIC-Universidad de Valencia, $^3$University of Graz}
}
\maketitle
\begin{abstract}
We examine the two solutions (massive and scaling) for the covariant Yang-Mills Dyson-Schwinger equations within stochastic quantization, and find that the scaling solution does not survive outside Landau gauge. 
We also see that the (rainbow) massive solution has less Faddeev-Popov effective action.
Finally, we argue that gluon confinement has only been marginally established in experiment and suggest further empirical work.
\end{abstract}

\section{The two solutions of YM theory within stochastic quantization}

\begin{figure}[htb]
\includegraphics*[width=0.47\textwidth]{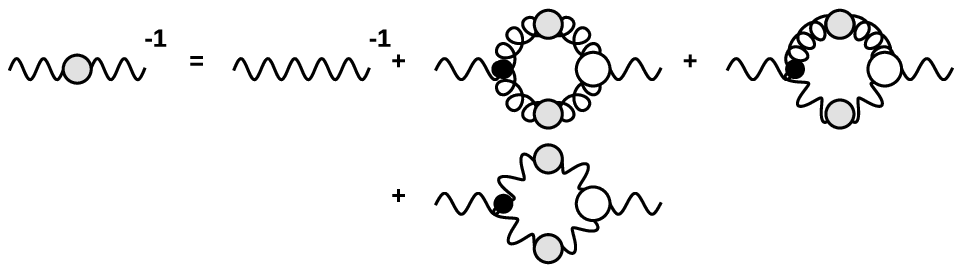}
\hspace{0.3cm}  
\includegraphics*[width=0.47\textwidth]{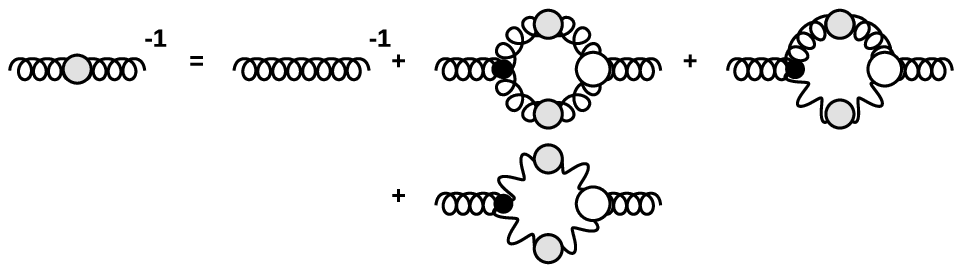}\\
\includegraphics*[width=0.5\textwidth]{FIGS.DIR/gluonL.eps}
\includegraphics*[width=0.5\textwidth]{FIGS.DIR/gluonT.eps}\\
\includegraphics*[width=0.5\textwidth]{FIGS.DIR/gluonLa.eps}
\includegraphics*[width=0.5\textwidth]{FIGS.DIR/gluonTa.eps}
\caption{Longitudinal (left panels) and transverse (right panels) gluon propagator dressing functions $Z_L$ and $Z_T$ in stochastic quantization. 
The top row displays the respective Dyson-Schwinger equations in rainbow approximation.
The middle row shows both the scaling and the massive (also called decoupling) solutions in Landau gauge $a=0$.  The bottom row is obtained with a small but finite $a$ parameter and shows the massive solution only; the scaling solution ceases to exist for $a\ne 0$.}
\label{fig:stochastic}
\end{figure}

Two families of solutions to the wave equations of Yang-Mills theory have been widely studied. The first is called ``massive'' or ``decoupling''~\cite{Dudal:2007cw} and characterized by a dynamically generated gluon mass scale, with Euclidean gluon propagator $Z(k^2)/k^2\propto 1/(k^2+m^2)$. The second is a ghost dominated, gluon suppressed ``scaling'' solution~\cite{Alkofer:2003jr} with respective power-law behaviors $Z(k^2)/k^2\propto (k^2)^{2\kappa-1}$ and $G(k^2)/k^2 \propto (k^2)^{-\kappa-1}$ (with $\kappa>0.5$).  Lattice gauge theory finds the massive solution in Landau-gauge fixed simulations~\cite{Cucchieri:2009zt}. Much discussion has focused on the ability to fix the gauge in large lattices and the presence of Gribov copies.

To continue studying the impact of the formation of a Gribov horizon and the gauge dependence of the solutions~\cite{Zwanziger:1988jt} we have performed~\cite{LlanesEstrada:2012my} a numerical analysis within stochastic quantization~\cite{Parisi:1980ys}.
In this approach to Yang-Mills theory the weight employed to compute correlators 
$
\la A(x) A(y) \ra = \int DA A(x) A(y) e^{-S_{\rm (YM)}[A]}\ ,
$
akin to a Boltzmann equilibrium distribution $e^{-\beta E}$,
is seen as the end-point $e^{-S_{\rm (YM)}[A]}=\lim_{\tau\to\infty}
P(\tau)$ of a stochastic random walk in a fictitious time parameter $\tau$. $P$ satisfies a Fokker-Planck equation with force $K^a_\mu(x)\equiv-\frac{\delta S_{\rm (YM)}}{\delta A^a_\mu(x)}$,
$$
\frac{\partial P}{\partial \tau} = \int d^4x \frac{\delta}{\delta A^{a \mu}(x)}
\left(
\frac{\delta P}{\delta A^a_\mu(x)} - K^a_\mu(x) P
\right) \ .
$$
To avoid the stochastic evolution running away along a gauge orbit (line of constant action), Zwanziger added a force that respects gauge-independent dynamics, a  gauge transformation
$
K^a_\mu(x)\to -\frac{\delta S_{\rm (YM)}}{\delta A^a_\mu(x)} + a^{-1} D^{ac}_{\mu}\partial\cdot A^{c}(x).
$
The real constant $a$ controls the relative intensity of the stochastic Yang-Mills and the gauge-restoring forces. The gauge is not strictly fixed, rather, gauge-equivalent configurations are weighted in a smooth manner, with much less probability for those farther from $A=0$, except in the limit $a\to 0$ that fixes the Landau gauge.
The Gribov problem is thus bypassed.
While $P(\infty)$ is not known, its uniqueness and positivity, as well as the Dyson-Schwinger equations have been derived. Since this ``soft'' gauge fixing method uses no Faddeev-Popov ghosts, one has instead both transverse and longitudinal
dressing functions of the gluon propagator,
$$
\int d^4 x \la A^a_\mu(0) A^b_\nu(x)\ra e^{ik\cdot x} 
=\delta^{ab} \left( \frac{Z_T(k^2)}{k^2} \left(\delta^{\mu\nu} -\frac{k^\mu k^\nu}{k^2} \right) 
 + a \frac{Z_L(k^2)}{k^2} \frac{k^\mu k^\nu}{k^2} \right) \ .
$$
Solving the rainbow DSE's for $Z_T$ and $Z_L$ (see figure~\ref{fig:stochastic}) we find that the scaling solution can be found only in Landau gauge, not for finite $a$, suggesting indeed a connection to the Gribov horizon, while massive solutions can be found for both $a=0$ and finite $a$.

\section{Effective action in Faddeev-Popov formalism}

\begin{figure}[htb]
\centering
\begin{minipage}{0.15\textwidth}
\includegraphics*[width=0.85\textwidth]{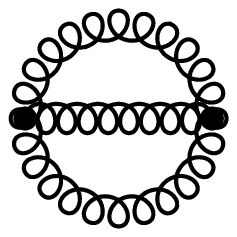}\\
\includegraphics*[width=0.85\textwidth]{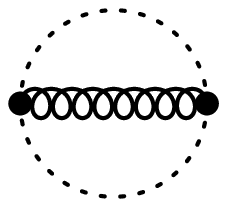}
\end{minipage}
\begin{minipage}{0.5\textwidth}
\includegraphics*[width=\textwidth]{FIGS.DIR/action.eps}
\end{minipage}
\caption{(left) Interaction terms of the Faddeev-Popov effective action generating the rainbow-DSE for the Yang-Mills propagators, and (right) evaluation of the action; $\alpha=0$ corresponds to a bare propagator, $\alpha=1/2$ to the massive/decoupling DSE solution, and $\alpha=1$ to the scaling propagators. The solid line interpolates between them and has a minimum at the massive solution. The dashed and dashed-dotted lines separately inform of the free and the interacting parts of the action.}
\label{fig:action}
\end{figure}

To understand from the continuum DSE perspective why lattice data favors the massive-like solutions we have examined the effective action~\cite{Berges:2004pu} $\Gamma[D,G]$ that generates DSE equations via $\delta \Gamma/\delta D =0$, $\delta \Gamma/\delta G=0$. We have evaluated the effective action for the bare (perturbative) propagators, for the massive propagators, and for the scaling ones. For simplicity we have limited ourselves to the rainbow DSE's in Faddeev-Popov formalism

The outcome, reported in figure~\ref{fig:action}, clearly shows that the massive propagator has least action in an unconstrained minimization, with the scaling solution disfavored. A natural question to ask is whether a constrained minimization can pick up the scaling solution in DSE or in a lattice computation, and whether such constraint is necessary from the point of view of the Gribov horizon formation in Landau gauge or similar considerations.

\section{Empirical studies of gluon confinement}

\begin{figure}[htb]
\centerline{\includegraphics*[width=0.5\textwidth]{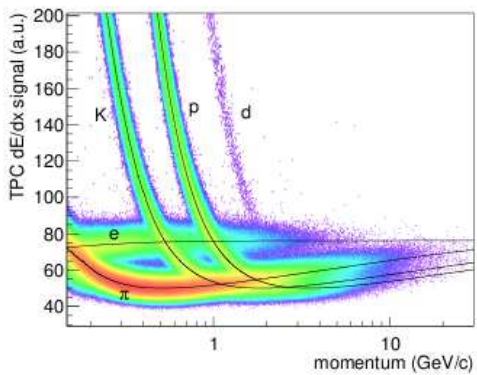}}
\caption{Energy deposition per unit length for charged particles in a detector. Particles with different $q/m$  are identified as bands. Free quarks are excluded, there not being a band directly left of the kaon's. Figure courtesy of P. Ladron de Guevara.}
\label{fig:dEdx}
\end{figure}

Searching for empirical evidence for either Yang-Mills scenario made us observe that not even gluon confinement stands on solid empirical footing~\cite{HidalgoDuque:2011je}.
What has been experimentally established is \emph{quark} confinement, by modern reassesments of Millikan's experiments against fractional charges at rest~\cite{Perl:2009zz} as well as energy deposition in high energy reactions~\cite{Bergsma:1984yn} that exclude energetic quarks (see figure~\ref{fig:dEdx}).

Possible evidence for gluon confinement could come from meson decays,
since $\pi_0\to \gamma \gamma$ accounts for the $\pi_0$ width, not leaving room for 
the gluon reaction $\pi_0\to g g$. 
The decay is however kinematically closed for both infrared solutions, that are gapped. $\Upsilon\to ggg$ has sufficient phase space, but 40 keV of its width is currently unaccounted for, so only a loose bound $\sigma_{b\bar{b}}\to ggg< 0.1\mu$barn can be obtained~\cite{HidalgoDuque:2011je}.

\begin{figure}[htb]
\centerline{\includegraphics*[width=0.5\textwidth]{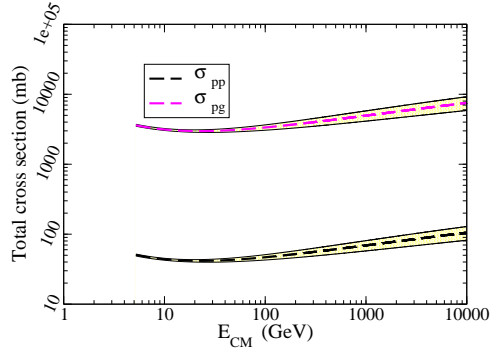}}
\caption{Gluon-proton cross section (top band) for high energy gluons obtained from Regge theory, with input the total proton-proton cross section (bottom band) and a rescaling of the pomeron coupling based on color counting alone.}
\label{fig:secgluonproton}
\end{figure}

\begin{figure}[htb]
\includegraphics[width=0.5\textwidth]{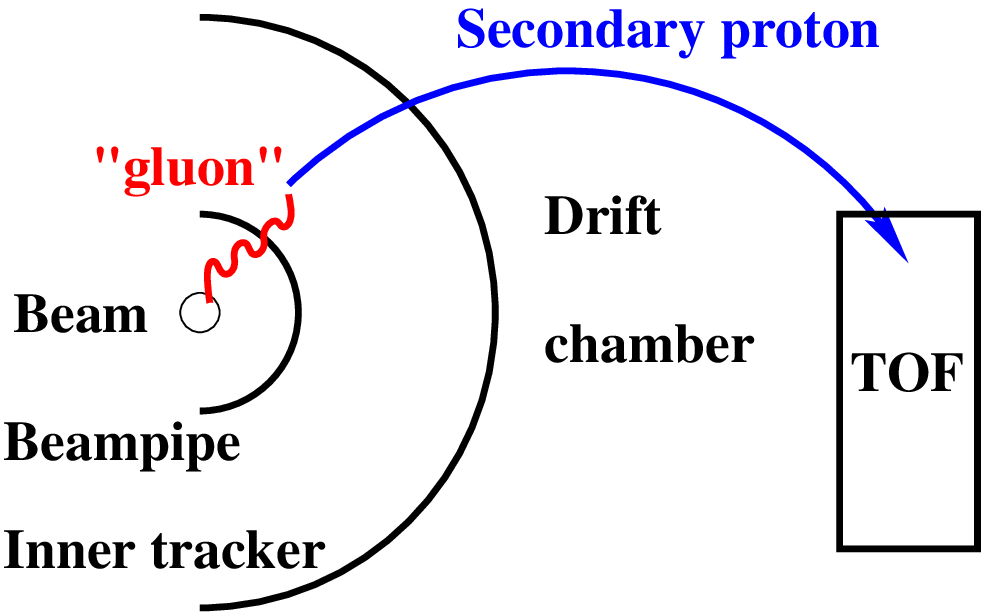}
\includegraphics[width=0.5\textwidth]{FIGS.DIR/exppenetration.eps}
\caption{Proposal to obtain collider bounds on gluon confinement. If the would-be liberated gluon reached the detector material, it would interact very strongly with atomic nuclei and eject a secondary proton (left), identified by its energy deposition and time-of-flight detector signal. The neutron background at the mbarn level can be ameleorated by noting that secondary protons would be spallated much earlier by gluons than they are by neutrons (right). Gluon liberation bounds at the $\mu$barn to nbarn level, depending on gluon energy, should be expected in pp collisions.}
\label{fig:gdetection}
\end{figure}

We suggest to further constrain free gluon production at hadron colliders such as the LHC. If a liberated gluon would reach the beampipe or vertex detector, because of its color charge, it would have a very short mean free path of about 0.6cm (see figure~\ref{fig:secgluonproton} for an estimate of the cross section based on Regge physics and large $N_c$~\cite{Brodsky:1973hm}). A sketch of our experimental proposal is detailed in figure~\ref{fig:gdetection}.

\section{Conclusion}

We have recalled the two widely studied behaviors for the infrared gluon. A massive-like propagator seems favored by lattice data in Landau gauge and we have shown that the effective action in the DSE formalism also suggests that this massive solution appears under unconstrained minimization. \\
The scaling solution where the Green's functions are all power laws remains nevertheless of theoretical interest, and we have shown that in stochastic quantization it appears only in Landau gauge, so perhaps it is a feature of Gribov horizon formation.\\
Although exploring alternative behaviors of IR glue in experiment seems appealing, we have noticed that gluon confinement itself is not on solid empirical footing, and have suggested to set bounds on it at hadron colliders.



\begin{thebibliography}{50}


\bibitem{Dudal:2007cw}
  D.~Dudal {\it et al.}
  Phys.\ Rev.\ D {\bf 77} (2008) 071501;
  A.~C.~Aguilar, D.~Binosi and J.~Papavassiliou,
  Phys.\ Rev.\ D {\bf 78} (2008) 025010;
  {\it ibid.} 
  JHEP {\bf 1201} (2012) 050;
  P.~Boucaud {\it et al.}
  JHEP {\bf 0806} (2008) 099.


\bibitem{Alkofer:2003jr}
  R.~Alkofer {\it et al.},
  Phys.\ Rev.\ D {\bf 68} (2003) 045003;
  R.~Alkofer, C.~S.~Fischer and F.~J.~Llanes-Estrada,
  Phys.\ Lett.\ B {\bf 611} (2005) 279
   [Erratum-ibid.\  {\bf 670} (2009) 460];
  C.~S.~Fischer and J.~M.~Pawlowski,
  Phys.\ Rev.\ D {\bf 75} (2007) 025012;
  D.~Zwanziger,
  Phys.\ Rev.\ D {\bf 65} (2002) 094039.






\bibitem{Cucchieri:2009zt}
  A.~Cucchieri and T.~Mendes,
  Phys.\ Rev.\ D {\bf 81} (2010) 016005;
  E.~-M.~Ilgenfritz {\it et al.}
  Phys.\ Rev.\ D {\bf 83} (2011) 054506;
  O.~Oliveira and P.~Bicudo,
  J.\ Phys.\ G {\bf 38} (2011) 045003.

\bibitem{Zwanziger:1988jt}
  D.~Zwanziger,
  Nucl.\ Phys.\ B {\bf 321} (1989) 591;
  C.~S.~Fischer and D.~Zwanziger,
  Phys.\ Rev.\ D {\bf 72} (2005) 054005;
  N.~Vandersickel and D.~Zwanziger,
  arXiv:1202.1491 [hep-th].



\bibitem{LlanesEstrada:2012my}
  F.~J.~Llanes-Estrada and R.~Williams,
  Phys.\ Rev.\ D {\bf 86} (2012) 065034.




\bibitem{Parisi:1980ys}
  G.Parisi and Y.-s.Wu,
  Sci.Sin.  {\bf 24} (1981) 483;
  D.~Zwanziger,
  Phys.\ Rev.\ D {\bf 67} (2003) 105001;
  R.F.~Alvarez-Estrada and A.~Munoz Sudupe,
  Phys.\ Rev.\ D {\bf 37} (1988) 2340;
  A.~Nakamura,
  Mod.\ Phys.\ Lett.\ A {\bf 6} (1991) 3331;
  J.M.~Pawlowski, D.~Spielmann and I.-O.~Stamatescu,
  Nucl.\ Phys.\ B {\bf 830} (2010) 291.

\bibitem{Berges:2004pu}
  J.~Berges,
  Phys.\ Rev.\ D {\bf 70} (2004) 105010.

\bibitem{HidalgoDuque:2011je}
  R. L. Delgado, C.~Hidalgo-Duque and F.~J.~Llanes-Estrada,
  arXiv:1106.2462 [hep-ph] (to appear in Few Body Physics, in press).

\bibitem{Perl:2009zz}
  M.~L.~Perl, E.~R.~Lee and D.~Loomba,
  Ann.\ Rev.\ Nucl.\ Part.\ Sci.\  {\bf 59} (2009) 47;
  M.~L.~Perl, E.~R.~Lee and D.~Loomba,
  Mod.\ Phys.\ Lett.\ A {\bf 19} (2004) 2595.


\bibitem{Bergsma:1984yn}
  F.~Bergsma {\it et al.}  [CHARM Collaboration],
  Z.\ Phys.\ C {\bf 24} (1984) 217;
  M.~L.~Stevenson,
  Phys.\ Rev.\ D {\bf 20} (1979) 82;
  D.~Antreasyan  {\it et al.},
  Phys.\ Rev.\ Lett.\  {\bf 39} (1977) 513;
  T.~Nash {\it et al.}
  Phys.\ Rev.\ Lett.\  {\bf 32} (1974) 858.


\bibitem{Brodsky:1973hm}
  S.J.~Brodsky, F.E.~Close and J.F.~Gunion,
  Phys.\ Rev.\ D {\bf 8} (1973) 3678;
  P.V.~Landshoff, J.C.~Polkinghorne, R.D.~Short,
  Nucl.\ Phys.\ B {\bf 28} (1971) 225.
  F.~J.~Llanes-Estrada {\it et al.}
  Nucl.\ Phys.\ A {\bf 710} (2002) 45;
  Y.~A.~Simonov,
  Phys.\ Lett.\ B {\bf 249} (1990) 514;
  J.~R.~Pelaez,
  Int.\ J.\ Mod.\ Phys.\ A {\bf 20} (2005) 628;
  J.~R.~Pelaez and F.~J.~Yndurain,
  Phys.\ Rev.\ D {\bf 69} (2004) 114001.


\end{thebibliography}
\end{document}